\documentclass{ws-ijbc}
\usepackage{ws-rotating}     % used only when sideways tables/figures are used
\begin{document}

\catchline{}{}{}{}{} % Publisher's Area please ignore

\markboth{N. Delis, C. Efthymiopoulos and G. Contopoulos} {Quantum
vortices and trajectories in particle diffraction}

\title{QUANTUM VORTICES AND TRAJECTORIES\\ IN PARTICLE DIFFRACTION}

\author{N. Delis}
\address{Research Center for Astronomy, Academy of Athens\\
Soranou Efesiou 4, Athens, GR-11527, Greece, and\\
Section of Astrophysics, Astronomy, and Mechanics, Department of
Physics,\\University of Athens,Panepistimiopolis Zografos, Athens,
GR15783, Greece\\nikdelis@sch.gr}
\author{C. Efthymiopoulos}
\address{Research Center for Astronomy, Academy of Athens\\
Soranou Efesiou 4, Athens, GR-11527, Greece\\
cefthim@academyofathens.gr}

\author{G. Contopoulos}
\address{Research Center for Astronomy, Academy of Athens\\
Soranou Efesiou 4, Athens, GR-11527, Greece\\
gcontop@academyofathens.gr}

\maketitle

\begin{history}
\received{(to be inserted by publisher)}
\end{history}

\begin{abstract}
We investigate the phenomenon of the diffraction of charged
particles by thin material targets using the method of the de
Broglie-Bohm quantum trajectories. The particle wave function can be
modeled as a sum of two terms $\psi=\psi_{ingoing}+\psi_{outgoing}$.
A thin separator exists between the domains of prevalence of the
ingoing and outgoing wavefunction terms. The structure of the
quantum-mechanical currents in the neighborhood of the separator
implies the formation of an array of \emph{quantum vortices}. The
flow structure around each vortex displays a characteristic pattern
called `nodal point - X point complex'. The X point gives rise to
stable and unstable manifolds. We find the scaling laws
characterizing a nodal point-X point complex by a local perturbation
theory around the nodal point. We then analyze the dynamical role of
vortices in the emergence of the diffraction pattern. In particular,
we demonstrate the abrupt deflections, along the direction of the
unstable manifold, of the quantum trajectories approaching an
X-point along its stable manifold. Theoretical results are compared
to numerical simulations of quantum trajectories. We finally
calculate the {\it times of flight} of particles following quantum
trajectories from the source to detectors placed at various
scattering angles $\theta$, and thereby propose an experimental test
of the de Broglie - Bohm formalism.
\end{abstract}

%\begin{multicols}{2}
\section{Introduction}
%%%%%%%%%%%%%%%%%%%%%%%%%%%%%%%%%%%%%%%%%%%%%%%%%%%%%%%%%%%%%%%%%%
The aim of the present paper is to implement the method of
\emph{quantum trajectories}, known as {\it de Broglie - Bohm} theory
\cite{debro1925,bohm1952} in the phenomenon of the diffraction of
charged particles by thin material targets. In the de Broglie - Bohm
theory we consider trajectories tracing the quantum currents. A
trajectory is determined by the initial particle's position and by
the pilot wave equation of motion
\begin{equation}\label{sch2}
{d\mathbf{r}\over dt}={\hbar\over
m}Im({\nabla\psi(\mathbf{r},t)\over \psi(\mathbf{r},t)})
\end{equation}
where $\psi$ is the wavefunction (`pilot wave'), $m$ is the particle
mass and $\hbar$ is Planck's constant. This equation implies
Newton's equation of motion in a potential
\begin{equation}\label{sch3}
{U(\mathbf{r},t)}=V(\mathbf{r},t)+Q(\mathbf{r},t)
\end{equation}
where ${Q(\mathbf{r},t)} $ is the `quantum potential', an extra term
caused by the wavefunction  $\psi$ :
\begin{equation}\label{sch4}
{Q(\mathbf{r},t)}=-{\hbar^2\over 2m}{\nabla^2|\psi|\over |\psi| }.
\end{equation}
The probability density of an ensemble of particles guided by the
same $\psi$ - field is
${\rho(\mathbf{r},t)}=|\psi(\mathbf{r},t)|^2$. The equations of
motion (\ref{sch2}) imply the continuity equation for $\rho$.
Furthermore, in the one-particle case, the Bohmian trajectories are
equivalent to the stream lines defined by the quantum probability
current $\mathbf{j}=(\hbar/2mi)
(\psi^*\nabla\psi-\psi\nabla\psi^*)$. Thus, the Bohmian method
yields practically equivalent results to Madelung's quantum
hydrodynamics \cite{mad1926}.

Bohm's trajectories have been proved useful in i) constructing
efficient numerical schemes for the integration of Schr\"{o}dinger's
equation via swarms of trajectories (e.g.
\cite{wya2005,sanzetal2002,ori2007}), and ii) visualizing processes
for which no other intuitive picture could be obtained. Examples of
the latter are the quantum tunneling effect \cite{lopwya1999}, the
(particle) two-slit experiment\cite{phietal1979}, ballistic
transport through `quantum wires' \cite{beehou1991}, molecular
dynamics \cite{gin2003}, dynamics in nonlinear systems with
classical focal points or caustics \cite{zhamak2003} and rotational
or atom-surface scattering \cite{sanzetal2004a}. A final application
regards the relation of chaotic quantum motions to the dynamical
origin of `quantum relaxation'
\cite{valwes2005,eftcon2006,ben2010,colstru2010}, namely the
justification of the approach (as time increases) of a quantum
system to Born's rule $\rho=|\psi|^2$, even if the initial
conditions were allowed to deviate from this rule
($\rho_{initial}\neq |\psi_{initial}|^2$).

In the case of the diffraction of charged particles, we find some
new phenomena due to the structure of quantum currents and to their
effects on quantum trajectories. In particular, we find that the
quantum flow is always characterized by the formation of an array of
{\it quantum vortices} delineated along a locus called {\it
separator}, i.e. a sharp boundary between the domains of prevalence
of the ingoing and outgoing flow. The local current structure in a
vortex forms a pattern called `nodal point - X-point complex'. This
has been previously investigated theoretically in
\cite{eftetal2007,coneft2008,eftetal2009}. Here we adapt this
analysis in the particular problem of particle diffraction, in order
to determine the number and location of all critical points of the
quantum flow, as well as to determine the size of the nodal point -
X-point complexes. The main result of this analysis is that the
deflection of quantum trajectories is due to their approaches close
to X-points. In fact, the trajectories follow the directions of the
asymptotic manifolds of the X-points, and this fact alone suffices
to explain the emergence of the whole diffraction pattern.

%_____________________________________________________________
%\begin{landscape}
\begin{figure}
\centering
\includegraphics[scale=0.50]{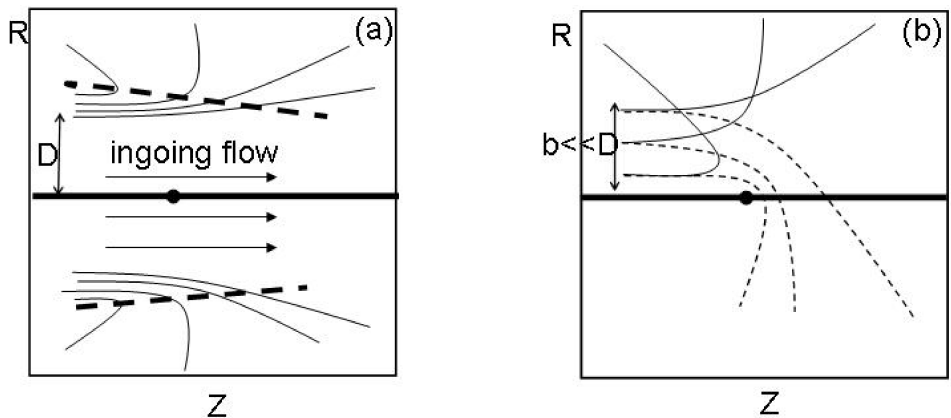}
\caption{(a) Schematic representation of the quantum trajectories in
the scattering or diffraction problem. Deflection takes place for
quantum trajectories with initial conditions of an $O(D)$ distance
away from the principal beam axis (z-axis). The thick dashed curves
illustrate the separator between ingoing and outgoing (from the
target) flow. The concentration of the quantum trajectories to
various Bragg angles is shown in Figure \ref{orbits} below. (b) Same
as in (a) but in the approximation of classical Rutherford
scattering for repelling (---), or attracting ($\cdot\cdot\cdot$)
forces. } \label{trjsch2}
\end{figure}
%\end{landscape}
%_____________________________________________________________
Passing to a global description of the quantum trajectories, we
identify a number of important differences between Bohmian and
classical trajectories, which are summarized with the help of a
schematic representation
(Figure \ref{trjsch2}). Four main differences are:\\

\noindent a) In the case of quantum trajectories, the  scattering
angle \emph{increases} when the impact parameter \emph{increases}
(Fig.1a)(this is explained in section 4), while the opposite is true
in the classical case.\\
b) The deflection of Bohmian trajectories takes place for initial
conditions at an $O(D)$ distance  from the principal axis of the
ingoing beam (z-axis), where $D$ is the transverse quantum coherence
length. A typical value for $D$ in electron beams is $D\geq
10^{-6}m$. On the contrary, for classical Rutherford scattering the
impact parameter $b$ is many orders
of magnitude smaller $(b\sim10^{-13} m$ ).\\
c) Bohmian  trajectories have the same form in both cases of
repelling and attractive forces (Figure \ref{trjsch2}a), a behavior
contrasted to the form of trajectories in the approximation of
classical Rutherford
scattering (Fig.\ref{trjsch2}b).\\
d) The most important difference regards the \emph{times of flight}
of particles to detectors placed at different scattering angles
$\theta$. The times of flight are shorter for larger impact
parameters in the quantum case, independently of the sign of the
interacting charges (while the times of flight depend on this sign
in the classical case and they are, in general, much shorter than in
the quantum case). This effect suggests an interesting experimental
test that will be described in section 5.

\section{Wave function}
We consider a cylindrical beam of particles of mass $m$ and charge
$Z_1e$ incident on a material target centered at the origin of the
coordinate system of reference. Cylindrical coordinates are denoted
by $z$ (horizontal), $R$ (transverse) and by the azimuth $\varphi$.
Use is also made of spherical coordinates $r=(z^2+R^2)^{1/2}$,
$\theta=\tan^{-1}(R/z)$, and $\varphi$.

The quantum-mechanical description of the diffraction process has
been discussed extensively (see, for example,
\cite{pengetal2004,peng2005} for reviews and further references).
Here, we adopt a simplified model which is sufficient for all
practical purposes in the analysis below. For the interaction of the
incident charged particles with any individual atom in the target,
we adopt a screened Coulomb potential
\begin{equation}\label{potatom}
U(\mathbf{r-r_j})={1\over 4\pi\epsilon_0}
{Z_1Ze^2\exp(-|\mathbf{r}-\mathbf{r}_j|/r_0)\over|\mathbf{r}-\mathbf{r}_j|}
\end{equation}
where $Z$ is the nuclear charge, $\mathbf{r}_j$ is the position of
the $j-th$ atom in the target and $r_0$ is a constant representing
the screening range, whose value is taken of the order of the atomic
size. The total potential felt by one incident particle is the sum
of the individual potentials:
\begin{equation}\label{pot}
V(\mathbf{r})=\sum_{j=1}^N U(\mathbf{r}-\mathbf{r}_j)~~
\end{equation}
where $N$ is the total number of atoms effectively participating in
the process of diffraction.

In order to describe diffraction quantum-mechanically, we specify a
model for the wavefunction $\psi$ of diffracted particles
(considering only elastic scattering). To this end, we first find
the eigenfunctions $\phi$ of the time-independent Schr\"{o}dinger
equation:
\begin{equation}\label{sch5}
-{\hbar^2\over 2m}\nabla^2 \phi + V(\mathbf{r})\phi = E\phi
\end{equation}
where $\phi$ is an eigenfunction corresponding to the energy value
$E>0$ of an incident particle, and $V(r)$ in (\ref{sch5}) is a
time-averaged form of (\ref{pot}). We solve Eq.(\ref{sch5})
following Born's approximation method (for $|V|<<E$)and expand
$\phi$ as
\begin{equation}\label{bornser} \phi=\phi_0+\phi_1+\phi_2+...
\end{equation}
where $\phi_0=O(1)$, $\phi_1=O(V/E)$, $\phi_2=O(V^2/E^2)$ etc. All
the essential phenomena appear already when only the two first terms
of the expansion $\phi\simeq \phi_0+\phi_1$ are considered. We then
find
\begin{equation}\label{phiall}
\phi_{\mathbf{k}}(\mathbf{r})\simeq e^{i\mathbf{k\cdot r}}
-{Z_1Ze^2\over 4\pi\epsilon_0}{m\over \hbar^2} {e^{ikr}\over r}
\left(\sum_{j=1}^Ne^{i\mathbf{(k-k_n)\cdot r_j}}\right) {1\over
k^2\sin^2(\theta/2)+1/r_0^2} \left[1+O\left({k<r_j^2>\over
r}\right)\right]
\end{equation}
where i) the label $\mathbf{k}$ refers to any vector in Fourier
space with modulus equal to $k=(2mE)^{1/2}/\hbar$, ii)
$\mathbf{k_n}=k\mathbf{n}$ with $\mathbf{n}=\mathbf{r}/r$, and iii)
$<r_j^2>=(1/N)\sum_{j=1}^N r_j^2$. The corrections due to the
$O(k<r_j^2>/r)$ and $1/r_0^2$ terms will be omitted since they do
not affect essentially the analysis.

The time-dependent eigenfunctions are now given by
$\psi_{\mathbf{k}}(\mathbf{r},t) =\exp(-i\hbar
k^2t/2m)\phi_{\mathbf{k}}(\mathbf{r})$. The electron wavefunction
can be modeled as a superposition of eigenfunctions
\begin{equation}\label{psiall}
\psi(\mathbf{r},t)={1\over (2\pi)^{3/2}} \int
d^3\mathbf{k}~\tilde{c}(\mathbf{k})\psi_{\mathbf{k}}(\mathbf{r},t)
\end{equation}
where $\tilde{c}(\mathbf{k})$ are Fourier coefficients. Because of
the collimation, the ingoing term of the wavefunction at $t=0$ can
be modeled as a traveling plane wave along the z-direction times a
Gaussian in the transverse direction with dispersion $\sim D$, i.e.
\begin{eqnarray}\label{psigauss}
\psi_{ingoing}(\mathbf{r},t=0)={1\over\sqrt{2}\pi D} \times
\exp\left(-{R^2\over 2D^2}+i(k_0 z) \right)~~,
\end{eqnarray}
where, assuming that the particles are monoenergetic, the constant
$k_0$ yields the average momentum $p_0=\hbar k_0$, or kinetic energy
$E_0=\hbar^2k_0^2/2m$ along the z-direction. The value of the
transverse quantum coherence length $D$, which turns to be a crucial
parameter in the analysis below, depends on the details of the
electron emission and collimation process.

The Fourier coefficients $\tilde{c}(\mathbf{k})$ for which
$\psi_{ingoing}$ has the form (\ref{psigauss}) are given by:
\begin{equation}\label{psifour}
\tilde{c}(\mathbf{k})=  \int
d^3\mathbf{r}~\psi_{ingoing}(\mathbf{r},t=0){e^{-i\mathbf{k\cdot
r}}\over (2\pi)^{3/2}}= {\delta(k_z-k_0)\over \pi^{1/2}\sigma_\perp}
\exp\left(-{k_x^2+k_y^2\over 2\sigma_\perp^2}\right)
\end{equation}
where $\sigma_\perp=D^{-1}$. Substituting (\ref{psifour}) into
(\ref{psiall}), and using (\ref{phiall}), we can evaluate the form
of the wavefunction at all times $t$. After some algebra we find
\begin{equation}\label{psiio}
\psi(\mathbf{r},t)=\psi_{ingoing}(\mathbf{r},t)+\psi_{outgoing}(\mathbf{r},t)
\end{equation}
where
\begin{eqnarray}\label{psiit}
\psi_{ingoing}(\mathbf{r},t)={1\over\sqrt{2}\pi}{D\over{\left(D^2+
i\hbar t/m\right)^{1/2}}} e^{-{R^2\over 2(D^2+i\hbar t/m)}+ i(k_0
z-k_0^2\hbar t/2m) }
\end{eqnarray}
\begin{eqnarray}\label{psiot}
\psi_{outgoing}(\mathbf{r},t)=-{1\over\sqrt{2}\pi} {D\over{\left(
D^2+ i\hbar t/m\right)^{1/2}}} {Z_1Ze^2\over
4\pi\epsilon_0}{m\over2\hbar^2} {S_{eff}(k_0;\theta,\varphi)\over
k_0^2\sin^2(\theta/2)}{e^{i(k_0r-\hbar k_0^2t/2m)}\over r} +
O\left({\sigma_\perp^2\over k_0^2}\right)
\end{eqnarray}
and $S_{eff}(k_0;\theta,\varphi)$, hereafter called the {\it
effective Fraunhofer function}, denotes the following sum over all
$N$ atomic positions $\mathbf{r}_j\equiv(x_j,y_j,z_j)$ in the
target:
\begin{eqnarray}\label{seff}
S_{eff}(k_0;\theta,\varphi)&=&\sum_{j=1}^N
\Bigg[\exp\bigg(-{x_j^2+y_j^2\over 2(D^2+i\hbar t/m)}\bigg)\nonumber\\
&\times&\exp\bigg(ik_0(2z_j\sin^2(\theta/2)
-x_j\sin\theta\cos\varphi -y_j\sin\theta\sin\varphi)\bigg)\Bigg]~~.
\end{eqnarray}
The effective Fraunhofer function $S_{eff}(k_0;\theta,\varphi)$
accounts for the usual diffraction effects. In numerical
calculations we consider, for simplicity, the case where the target
has a polycrystalline structure. Then, the diffraction pattern
becomes axisymmetric and the dependence of $S_{eff}$ on $\varphi$
disappears. We then derive a fitting model for $S_{eff}$ (see
Appendix A) which substitutes the sum (\ref{seff}) in all subsequent
calculations. This model reads:
\begin{eqnarray}\label{sbfin}
S_{eff}(\theta)&=&{D\over a}e^{i\delta}\Bigg[ \sum_{q=0}^{q_{max}}
C_{coherent}e^{-{1\over 2}4k_0^2\sin^4(\theta/2)\sigma_a^2}
\frac{2\sin\left[k_0d\sin(\theta_q)(\theta-\theta_q)/2\right]}
{k_0a\sin(\theta_q)(\theta-\theta_q)}\nonumber\\
&+&(1-e^{-{1\over
2}4k_0^2\sin^4(\theta/2)\sigma_a^2})C_{diffuse}\sqrt{d/a}\Bigg]
\end{eqnarray}
where $\theta_q$ are Bragg angles defined by
\begin{equation}\label{bragg}
\sin^2(\theta_q/2)={q\pi\over k_0a},~~~q=1,2,...q_{max}
\end{equation}
In Eq.(\ref{sbfin}), $d$ is the target thickness, $a$ is the
distance between nearest atoms (assuming for simplicity a cubic unit
cell, see appendix A). The first term in the r.h.s. of
Eq.(\ref{sbfin}) describes a coherent contribution to the outgoing
electron wave exhibiting sharp peaks at all Bragg angles $\theta_q$.
The second term describes diffuse scattering due e.g. to thermal
fluctuations or recoil effects in the target. These are modeled by
the constant $\sigma_a$ (dimension of length) which measures the
amplitude of random motions of the atoms (see Appendix A).
Accordingly, the exponential factors in Eq.(\ref{sbfin}) are
Debye-Waller factors. Finally, the constants $C_{coherent}$ and
$C_{diffuse}$ are fitting constants whose values are fixed by a
numerical simulation, while $\delta$ is an arbitrary phase (see
Appendix A).

In all subsequent calculations, we use the set of
Eqs.(\ref{psiit},\ref{psiot},\ref{sbfin}) which completely specify
the wavefunction. In numerical examples we adopt the following set
of parameter values: $Z_1=-1$, $m=m_e$, $k_0=8.877\times
10^2$nm$^{-1}$ (corresponding to electrons with energy $E=30$KeV, or
wavelength $\lambda_0=7\times 10^{-3}$nm), $D=1000$nm (corresponding
to a transverse quantum coherence length 1$\mu$m), $Z=79$ (gold),
$d=420$nm, $a=0.257$nm, $\sigma_a=0.0086$nm, $C_{coherent}=0.060$,
$C_{diffuse}=0.077$ (found by numerical fitting), $q_{max}=40$,
$\delta=\pi$. These values are relevant to the case of electron
diffraction, while they conform with a number of limitations imposed
by our approximations to the wavefunction model. Some main
limitations are:

a)The energy $E$ (determined by the particles' mass
$m$ and wavenumber $k_0$) must be well above the atomic energy levels
in the crystal (which are of the order of 1KeV) and below the nuclear
energies (which are of the order of 1MeV). In our case the energy
$E=30KeV$ is within these limits. Thus only the elastic term in the
outgoing wavefunction needs to be considered.

b) The use of Born's approximation is valid only sufficiently far
from all atoms so that $|V|<<E$. In the case of the potential of
Eqs.(\ref{potatom}) and (\ref{pot}), this is true at all points
$\mathbf{r}$ of space with distance greater than a few times
the size of the atoms in the target. All calculated Bohmian
trajectories in the sequel lie in regions exceeding by far such
distances.

c) The inequality $D^2>(\hbar t/m)$ sets an upper limit in the
time during which the ingoing wavefunction remains coherent in
the transverse direction. Setting $D=1\mu m$ as above, we find
$t<10^{-8}$sec, so that the total distance traveled by individual
electrons should be less than 1m, which is a quite long, hence a
feasible value.

Finally, we note that the representation of the beam as a plane wave
in the z-direction can also be improved by introducing a finite
`longitudinal coherence length' $l$, i.e. a wavepacket approach in
the z-direction as well. In this case, the fitting model of Appendix
A is no longer valid and a separate analysis must be made covering
the cases $D<l$ or $D>l$ (work in progress \cite{deletal2011}).

%%%%%%%%%%%%%%%%%%%%%%%%%%%%%%%%%%%%%%%%%%%%%%%%%%%%%%%%%%%%
\section{Quantum vortices}
%%%%%%%%%%%%%%%%%%%%%%%%%%%%%%%%%%%%%%%%%%%%%%%%%%%%%%%%%%%%

The form of the quantum currents in the model defined by Eqs.
(\ref{psiit},\ref{psiot},\ref{sbfin}) can be understood
qualitatively by the following remarks:\\

\subsection{Separator}
\begin{figure}[h]
\centering
\includegraphics[scale=0.60]{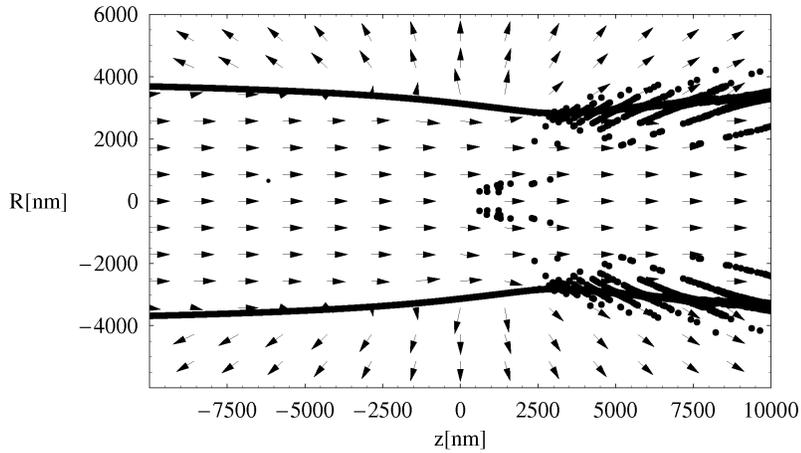}
\caption{Structure of the quantum currents in the model of
Eqs.(\ref{psiit},\ref{psiot},\ref{sbfin}) and parameters as
specified in the text.} \label{curgen}
\end{figure}

The ingoing and outgoing waves become equal in size if
$|\psi_{ingoing}| =|\psi_{outgoing}|$. Since $\hbar t/m<<D^2$ for
all times of interest, one has from Eqs.(\ref{psiit}) and
(\ref{psiot})
\begin{equation}\label{rthnod}
R\exp\left(-{R^2\over 2D^2}\right) \simeq G(\theta)\equiv \left(
{|Z_1Z|e^2\over 4\pi\epsilon_0}{m\over\ 2k_0^2\hbar^2}
{|S_{eff}(\theta)|\sin\theta\over \sin^2(\theta/2)}\right)~~~.
\end{equation}
The roots of Eq.(\ref{rthnod}) yield {\it a separator line} on the
meridian plane $(R,z=R/\tan\theta)$ which acts as a delimiter
between the domains of prevalence of the axial ingoing from the
radial outgoing flow. Figure \ref{curgen} corresponds to a numerical
calculation, where the arrows indicate the local direction of the
quantum flow at every point of the configuration space. Clearly, the
transition from axial ingoing to radial outgoing flow takes place
essentially through the separator lines (bold lines).

%_____________________________________________________________
%\end{landscape}
%_____________________________________________________________
Ray-like structures in the right part of Fig.(\ref{curgen})
correspond to the form taken by the separator locally near every
Bragg angle. Important details of the separator structure are not
discernible at this resolution, appearing only when we zoom very
close to one Bragg angle. From Eq.(\ref{rthnod}) we can see that
roots exist only if $G(\theta)$ satisfies $G(\theta)<C(D)$, where
$C(D)$ is the maximum value of $R\mbox{e}^{-{R^2/2D^2}}$, equal to
$D/\sqrt{\mbox{e}}$. Then the two roots $R_1(\theta)<D$ and
$R_2(\theta)>D$ define a lower and un upper branch of the separator.
Only the upper branch $R_2(\theta)$ is shown in Figure \ref{curgen},
since for most angles $\theta$ the lower branch corresponds to
values of $R_1(\theta)$ smaller than $R_2(\theta)$ by many orders of
magnitude.

%_____________________________________________________________
\begin{figure}[h]
\centering
\includegraphics[scale=0.50]{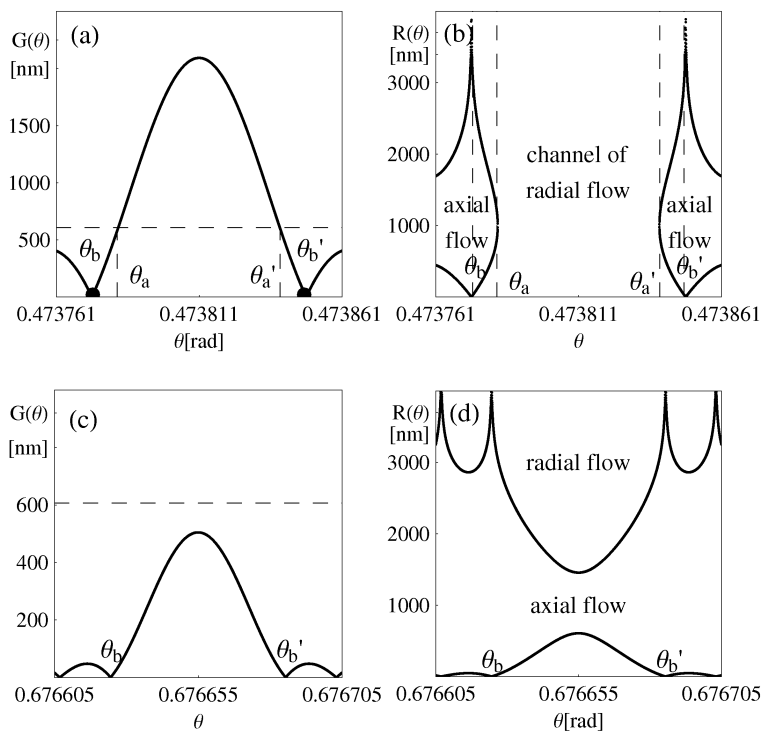}
\caption{(a) The function $G(\theta)$ in a small neighborhood of the
fourth Bragg angle $\theta_q=\theta_4=0.473811$rad. The horizontal
dashed line corresponds to the level value
$C(D)=De^{-0.5}=606.53$nm. The inner and outer separator curves are
joined at the angles $\theta_a=0.4737825$rad and
$\theta_a'=0.4738395$rad, and a channel of radial flow is formed, of
width $\Delta\theta=\theta_a'-\theta_a= 5.7\times 10^{-5}$rad. The
function $G(\theta)$ becomes zero at the two closest zeros of the
effective Fraunhofer function near the Bragg peak at
$\theta_q=\theta_4$, namely at the angles $\theta_b=0.473773$ and
$\theta_b'=0.47385$. At these angles $R_1$ becomes zero while $R_2$
tends to infinity. (b) The form of the separator curves $R(\theta)$
in the same range of angles $\theta$ as in (a). (c,d) Same as in
(a,b) but for a neighborhood of the Bragg angle
$\theta_q=\theta_8=0.676655$.} \label{channel}
\end{figure}
%_____________________________________________________________

The above picture changes very close to one Bragg angle $\theta_q$,
where $R_1(\theta)$ becomes also important. This is because the
local peaks of $|S_{eff}(\theta)|$ imply also local peaks of the
function $G(\theta)$ in Eq.(\ref{rthnod}). According to the local
value of the function $G(\theta=\theta_q)$ we distinguish the
following
two cases:\\

\noindent {\it Case I: $G(\theta_q)>C(D)$}. In this case there is an
interval of values $\theta_a<\theta<\theta_a'$ containing the Bragg
angle $\theta_q$ such that Eq.(\ref{rthnod}) has no roots within it
(Figure \ref{channel}a). Then, the inner separator $R_1(\theta)$ of
Fig.\ref{channel}b joins the outer separator $R_2(\theta)$ at the
angles $\theta_a$ and $\theta_a'$ and takes the form shown in
Fig.\ref{channel}b (in coordinates $R$ vs. $\theta$). The gap
between the left and right domains of prevalence of the ingoing flow
corresponds to a narrow angular strip (hereafter called a `channel')
along which the flow is radial.\\

\noindent {\it Case II: $G(\theta_q)<C(D)$}. In this case there are
roots of both $R_1(\theta)$ and $R_2(\theta)$ for all angles
$\theta$ surrounding and including $\theta_q$ (Fig.\ref{channel}c).
However, $R_1(\theta)$ has a local maximum and $R_2(\theta)$ a local
minimum at $\theta_q$ as well as on all other peaks caused by the
side lobes of the Fraunhofer function. The separator then develops
oscillations as shown in Fig.\ref{channel}d. Furthermore, in this
case there may be one or more pairs of angles $(\theta,\theta')$
where the inner, and outer, separator reaches the z-axis, and
infinity,
respectively. Two such pairs are visible in Fig.\ref{channel}d.\\

\subsection{Nodal point - X-point complexes}
Along the separator, a large number of quantum vortices are formed.
Their location is given by all points where the conditions i)
$\psi=0$, and ii)$\nabla\psi\neq 0$ hold. Such points are called
{\it nodal points}. The condition $\psi=0$, or $\psi_{ingoing}=
-\psi_{outgoing}$ yields, besides Eq.(\ref{rthnod}), a condition for
the equality of phases, which takes the form (ignoring again
$O(\hbar t/m)$ terms)
\begin{equation}\label{phaseeq}
k_0R\tan(\theta/2)=2\bar{q}\pi~~~~~~~~\bar{q}\in{\cal Z}~~.
\end{equation}
The pair of equations (\ref{rthnod},\ref{phaseeq}) specifies
completely the coordinates of one nodal point(R,$\theta$). We locate
many nodal points numerically by choosing different values of
$\bar{q}$. It follows from Eq.(\ref{phaseeq}) that the separation
between two nearby nodal points, defined by two values $\bar{q}$ and
$\bar{q}+1$, is of order $\pi/k_0\sim\lambda_0$.

%_____________________________________________________________
\begin{figure}
\centering
\includegraphics[scale=0.3]{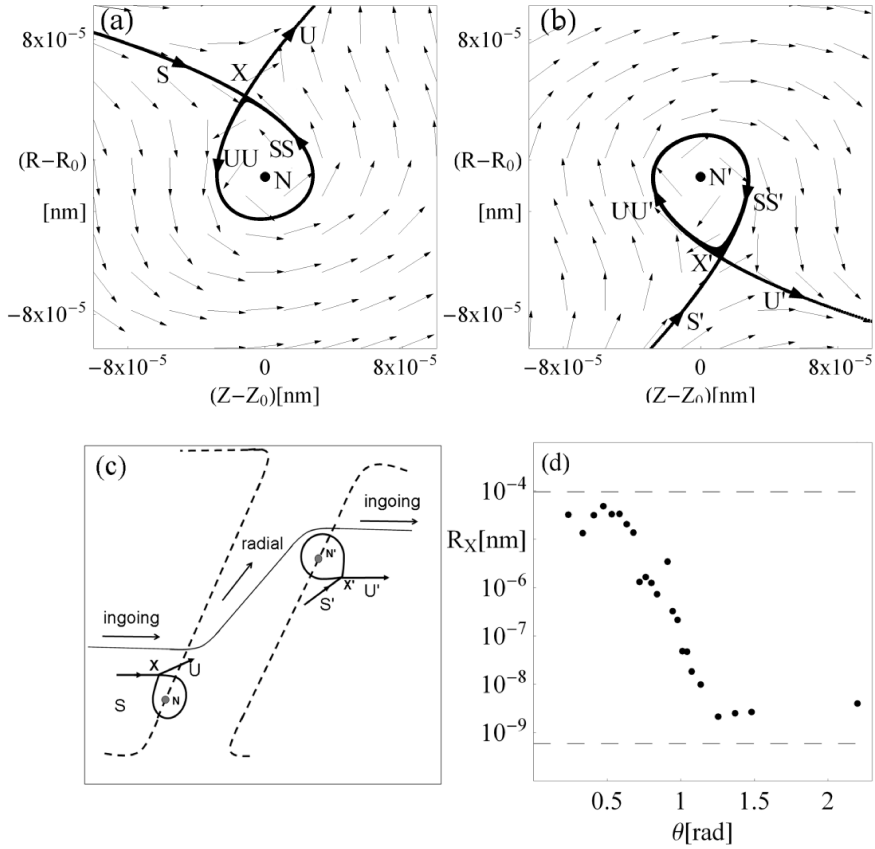}
\caption{(a) Instantaneous form of the quantum flow around a nodal
point placed at the left separator of the channel formed around the
Bragg angle $\theta_4$ ($z_0=4505.7354$nm, $R_0 = 2310.6028$nm). The
flow forms a `nodal point - X-point' complex (Nodal point (N),
X-point (X)) and follows in general one branch of the stable
manifold (S) and then one branch of the unstable manifold (U) of the
X-point. The other branches (SS,UU) are joined to form a loop. (b)
Same as in (a) but for a nodal point placed on the right separator
of the channel of $\theta_4$ ($z_0=4507.1939$, $R_0=2310.9549$nm).
(c) Schematic representation of the quantum flow at the crossing of
the channel, along with the directions of the stable and unstable
manifolds of the X-points formed near the separator nodal points.
The dashed lines represent the separator. (d) The size of some nodal
point - X-point complexes, quantified by the distance $R_X$ from the
nodal point to the X-point, as a function of $\theta$. The upper and
lower dashed lines correspond to the estimates of Eq.(\ref{rxest})
for the domains of Bragg angles and of diffuse scattering
respectively.} \label{nodal}
\end{figure}
%_____________________________________________________________
The local form of the quantum flow in a very small neighborhood
around one nodal point is very different from the general picture of
the flow as given in Fig.\ref{curgen}. If we `freeze' the time $t$,
the instantaneous pattern formed by the vector field of quantum
probability current $\mathbf{j}$ corresponds to a characteristic
structure called {\it nodal point - X-point complex}
\cite{eftetal2007,coneft2008,eftetal2009}. That is, close to a nodal
point we find a second critical point of the flow, where one has
$\mathbf{j}=0$. This is called an `X-point', since it can be shown
that it is always simply unstable, i.e. there are two real
eigenvalues of the matrix of the linearized flow around X, which are
one positive and one negative. Accordingly, there are two opposite
branches of unstable (U) and stable (S) manifolds emanating from X.
On the other hand, the nodal point can be an attractor, center, or
repellor. This determines the local form of the invariant manifolds
U and S. It has been established theoretically \cite{eftetal2007}
that, except for a set of very small measure, most quantum
trajectories {\it avoid} the nodal point, being instead scattered
along the asymptotic directions of the manifolds of the X-point,
leading to large distances from the nodal point - X-point complex.
Furthermore, while, in general, the motion of nodal point - X-point
complexes introduces chaos
(\cite{fri1997,wispuj2005,eftetal2007,eftetal2009}), in the present
problem this effect is negligible because i) the speed of vortices
is extremely small (of order $\sim\hbar/(k_0mD^2)<<v_0$), and ii)
the quantum trajectories exhibit only a finite number of encounters
with nodal point - X-point complexes, as will be shown with
numerical examples below. In conclusion, the effect of the nodal
point - X-point complexes on the trajectories can be described as a
scattering process without recurrences.

Figures \ref{nodal}a,b show two examples of such nodal point -
X-point complexes, in which the central nodal points are located on
the left side (Fig.\ref{nodal}a) and right side (Fig.\ref{nodal}b)
respectively of the main channel around the fourth Bragg angle shown
in Fig.\ref{channel}b. The stable and unstable manifolds of the
X-points (found numerically) are shown like thick curves in the same
figures. In Fig.\ref{nodal}a, the left branch of the stable manifold
is almost horizontal far from the X-point, while the right branch of
the unstable manifold follows a radial direction which coincides
with the direction defined by the local value of the scattering
angle $\theta$. The other two branches form loops around the nodal
point. If the terms $\hbar t/m$ in the wavefunction are ignored, the
two branches join each other smoothly and the nodal point is a
center. If however, all terms are taken into account, the nodal
point is either an attractor or a repellor (see \cite{eftetal2009}).
Then one of the two branches in-spirals towards the nodal point,
while the other deviates outwards after forming a nearly complete
loop around the nodal point. In the present case, the time
dependence of the wavefunction introduces negligible effects and the
loops formed around all nodal points can be considered as
practically closed. Then, in Fig.\ref{nodal}a we see that, due to
the form of the stable and unstable manifolds, trajectories
approaching the X-point from the left in a nearly horizontal
direction are scattered by the X-point. A scattered trajectory may
or may not form a loop around the nodal point. In either case, the
trajectories eventually recede from the X-point along the unstable
manifold, and follow asymptotically a radial direction (upwards and
to the right in Fig.\ref{nodal}a). However, these directions are
swapped in Fig.\ref{nodal}b. The swapping can be understood with the
help of a schematic figure (Fig.\ref{nodal}c).

Details on the size of a nodal point - X-point complex are found by
expanding the wavefunction $\psi$ as well as the Bohmian equations
of motion in variables $u=z-z_0$, $v=R-R_0$ around a nodal point
$(z_0,R_0)$.

Namely, from Eqs.(\ref{psiit},\ref{psiot}), the wavefunction takes
the form
\begin{eqnarray}\label{psialla2}
\psi&=&\left[{1\over\sqrt{2}\pi}{D\over{\left(D^2+ i\hbar
t/m\right)^{1/2}}} e^{-ik_0^2\hbar t/2m}\right]\times\psi'
\end{eqnarray}
where
\begin{equation}\label{psipa2}
\psi'= e^{-{R^2\over 2(D^2+i\hbar t/m)}+ik_0 z }
-{\bar{P}S_{eff}(\theta)\over \sin^2(\theta/2)}{e^{ik_0r}\over r}
\end{equation}
and $\bar{P}=(Z_1Ze^2m)/(8\pi\epsilon_0\hbar^2)$. The prefactor in
front of $\psi'$ in Eq.(\ref{psialla2}) is simplified in the Bohmian
equations of motion, which take the form
\begin{equation}\label{eqmoa2}
{dz\over dt}= {\psi'_*(\partial\psi'/\partial
z)-\psi'(\partial\psi'_*/\partial z) \over\psi'\psi'_*},~~~~
{dR\over dt}= {\psi'_*(\partial\psi'/\partial
R)-\psi'(\partial\psi'_*/\partial R) \over\psi'\psi'_*}~~.
\end{equation}
Following \cite{eftetal2009}, the main characteristics of the
equations of motion are found by the second order development of
$\psi'$ around a nodal point $(z_0,R_0)$. For $\bar{P}S_{eff}>0$ (as
in the numerical parameters above), the condition $\psi'(z_0,R_0)=0$
yields, if we disregard the term $i\hbar t/m$,
$$
\bar{P}S_{eff}(\theta_0)={R_0^2e^{-R_0^2/2D^2}\over 2(r_0+z_0)},~~~
e^{ik_0z_0}=e^{ik_0r_0}
$$
where $r_0=\sqrt{R_0^2+z_0^2}$ and $\theta_0=\tan^{-1}(R_0/z_0)$.
Inserting this expression into the second order development of
$\psi'$,
\begin{eqnarray}\label{psiexp}
\psi'(u,v;z_0,R_0)&=&(a_{10}+ib_{10})u+(a_{01}+ib_{01})v\nonumber\\
&+&{1\over 2}(a_{20}+ib_{20})u^2+{1\over 2}(a_{02}+ib_{02})v^2
+(a_{11}+ib_{11})uv+\ldots,
\end{eqnarray}
where $u=z-z_0$, $v=R-R_0$, we find expressions for the coefficients
$a_{10}$, $b_{10}$ ,$a_{01}$ ,$b_{01}$, $a_{20}$ ,$b_{20}$,
$a_{02}$, $b_{02}$, $a_{11}$, $b_{11}$ (see Appendix B). The
position of the X-point in the `adiabatic approximation' (i.e.
ignoring time variations of $\psi'$) can be estimated (see equations
(4) and (11) of \cite{eftetal2009}) as:
\begin{eqnarray}\label{xpointa2}
0\simeq Av_X+B_1u_X^2+C_1v_X^2+D_1u_Xv_X\nonumber\\
0\simeq -Au_X+B_2u_X^2+C_2v_X^2+D_2u_Xv_X
\end{eqnarray}
where
$$
A=a_{01}b_{10}-a_{10}b_{01},
$$
$$
B_1={a_{02}b_{10} - a_{10}b_{02}\over 2},~~ C_1={a_{02}b_{10} -
a_{10}b_{02} - 2a_{11}b_{01}\over 2},~~ D_1=a_{02}b_{01} -
a_{01}b_{02}
$$
$$
B_2={a_{01}b_{02} - a_{02}b_{01}\over 2},~~ C_2={a_{01}b_{02} -
a_{02}b_{01} - 2a_{11}b_{10}\over 2},~~ D_2=a_{10}b_{02} -
a_{02}b_{10}~~.
$$
The variational matrix at $(u_X,v_X)$ yielding the eigenvalues and
eigenvectors at the X-point is determined by the same coefficients.

In Appendix B we estimate the size of all coefficients
$a_{ij},b_{ij}$ near and far from Bragg angles. Using the set of
equations (\ref{xpointa2}) we then find an estimate for the quantity
$R_X=(u_X^2+v_X^2)^{1/2}$, i.e. for the size of a nodal point -
X-point complex. The final result is
\begin{eqnarray}\label{rxest}
R_X=O\left({d\over Dk_0}\right)~~&\mbox{domain of Bragg angles},~~~~\\
R_X=O\left({1\over Dk_0^2}\right)~~&\mbox{domain of diffuse
scattering}~.\nonumber
\end{eqnarray}
Figure \ref{nodal}d shows the dependence of $R_X$ on $\theta$,
obtained by computing numerically a sample of nodal point -X-point
complexes along the separator at various angles $\theta$. The upper
and lower dashed lines correspond to the upper and lower estimates
of Eq.(\ref{rxest}) respectively. We note that the transition from
the validity of one estimate to the other occurs around the angle
$\theta\approx 0.8$rad, This is nearly the angle beyond which the
second term in the r.h.s. of Eq.(\ref{sbfin}) (diffuse scattering)
becomes more important than the first term (coherent scattering).

%%%%%%%%%%%%%%%%%%%%%%%%%%%%%%%%%%%%%%%%%%%%%%%%%%%%%%%%%%%%%%%
\section{Quantum trajectories. Emergence of the diffraction pattern}
%%%%%%%%%%%%%%%%%%%%%%%%%%%%%%%%%%%%%%%%%%%%%%%$$$$$$$$$$$$$$$%

\subsection{Trajectories}
\begin{figure}
\centering
\includegraphics[scale=0.55]{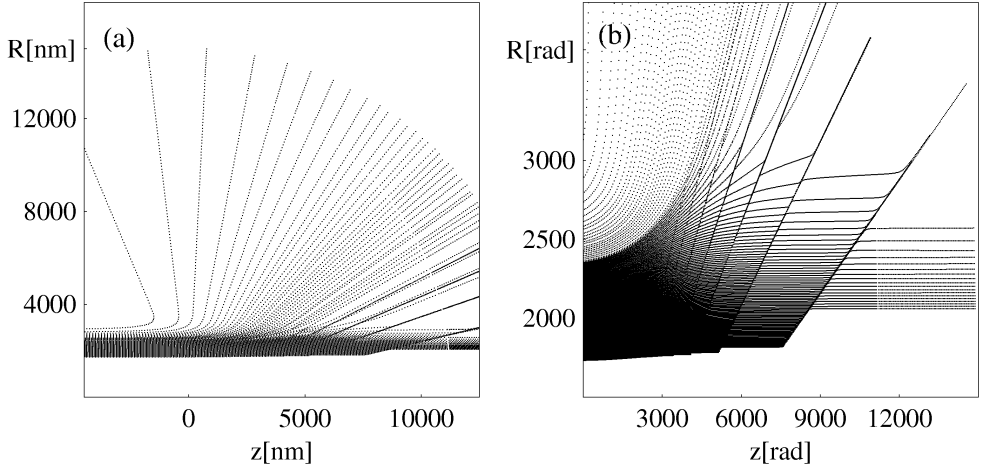}
\caption{(a) A swarm of quantum trajectories (initial conditions:
$z_0=-10^4$nm, $R_0$ in the interval $1.5\times 10^3$nm$\leq R_0\leq
3.3\times 10^3$nm). The plot shows a sample of trajectories selected
by their values of $R_0$ being distributed according to ${\Delta
N\over \Delta R}\propto 2\pi R \mbox{e}^{-R^2/D^2}$ which
corresponds to the choice of $\psi_{ingoing}$ as in
Eq.(\ref{psiit}). (b) A zoom of (a) in the region where the
trajectories are forced to follow a diffraction pattern by crossing
the channels of consecutive Bragg angles. The sharp deflections seen
correspond to the Bragg angles $\theta_q$, $q=1,...,8$ (right to
left).} \label{orbits}
\end{figure}

The results of the previous section will now be used in order to
understand the form of Bohm's trajectories of diffracted particles.
A numerical calculation of a swarm of such trajectories is shown in
Figure \ref{orbits}a. The initial conditions were chosen as
$z_0=-10\mu$m, and $x_0$ taken uniformly in the interval
$[1.5,3.3]\mu$m, for 360 trajectories.  In Fig.\ref{orbits}a a
selective sample of trajectories is plotted so as to follow, at
$t=0$ the distribution $dN/dR\propto R\mbox{e}^{-R^2/D^2}$, with
$D=1\mu$m, corresponding to the choice of $\psi_{ingoing}$  as in
Eq.(\ref{psiit}). The numerical integration was done with adjustable
time step of maximum value $\Delta t=0.1\pi\hbar/E$ (where
$E=30$KeV) so as to ensure that the segment of a trajectory covered
within one time step is significantly smaller than the wavelength
$\lambda_0$.

%_____________________________________________________________
%_____________________________________________________________
The main feature of Fig.\ref{orbits}a is that the trajectories with
larger initial normal distance $R_0$ from the z-axis are deflected
to larger angles $\theta$. This effect is in contrast to the picture
of classical Rutherford scattering and it can be regarded as a clear
manifestation of the quantum character of diffraction for a large
transverse quantum coherence length. Such behavior of the quantum
trajectories is readily accounted for by the fact that the average
inclination of the separator in the $(z,R)$ plane is negative.
Namely, since all trajectories are horizontal until they encounter
the separator, trajectories of larger initial $R_0$ encounter the
separator at a larger angle $\theta$ satisfying Eq.(\ref{rthnod})
with $R=R_0$(Fig.1a). We thus have that $\theta$ is an increasing
function of $R_0$.

\begin{figure}
\centering
\includegraphics[scale=0.55]{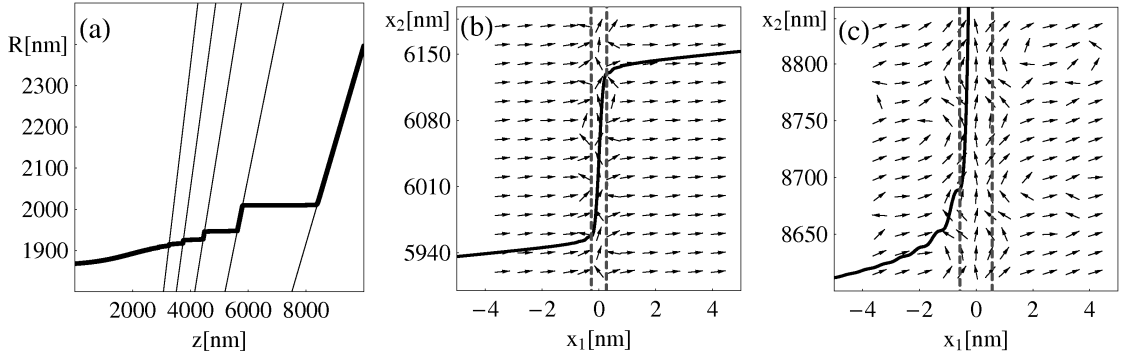}
\caption{(a) One quantum trajectory (bold), initial conditions
$(z_0,R_0)=(-10000$nm$,1865$nm$)$, which exits from the first Bragg
angle after a number of visible consecutive encounters with the
channels of subsequent Bragg angles (solid lines). (b) The crossing
of the channel of the second Bragg angle
$\theta_q=\theta_2=0.33345266$ as viewed in locally rotated
coordinates $x_1=z\cos(\pi/2-\theta_2)-R\sin(\pi/2-\theta_2)$,
$x_2=z\sin(\pi/2-\theta_2)+R\cos(\pi/2-\theta_2)$. Note the
different scale on the axes. The arrows indicate the local direction
of the quantum flow, which changes abruptly in a narrow zone near
$x_1=0$, corresponding to $\theta$ being exactly equal to the Bragg
value. The dashed vertical lines indicate the positions of the pair
of closest zeros to the peak of the effective Fraunhofer function
near $\theta=\theta_2$. (c) Same as in (b) but for the crossing of
the channel of the first Bragg angle $\theta_q=\theta_1=0.23523778$.
In this case the trajectory never reaches the right border of the
channel. } \label{orbdet}
\end{figure}

Figure \ref{orbits}b shows now a zoom in the domain of
forward-scattered trajectories, leading to the formation of a
diffraction pattern. The trajectories that started closer to the
z-axis follow initially the general flow due to $\psi_{ingoing}$,
but they are subject to abrupt deflections at the crossing of any
Bragg angle. Such deflections are due to the entering of the
trajectories into channels of radial flow. This effect is depicted
in detail in Figure \ref{orbdet}a, showing only one trajectory that
suffers five consecutive deflections by passing through consecutive
channels of radial flow before exiting to the domain of prevalence
of the radial flow leading to the first Bragg angle
$\theta_q=\theta_{1}=0.23$.  A comparison of the deflections at the
Bragg angles $\theta_q=\theta_{2}$ and $\theta_q=\theta_{1}$ is
shown in panels (b) and (c) of the same figure, where the
coordinates are locally rotated so that the vertical axis always
coincides with the radial direction. We clearly see the channels of
radial quantum flow formed around each Bragg angle. The dashed
vertical lines show local segments of the separator lines which mark
essentially the limits of the channels in either case. The
deflection of an orbit is caused at the crossing of the separator,
and it is due to the orbit necessarily following the flow around the
X-points that exist along the separator. The deflection causes a
trajectory to follow a path nearly parallel to the radial direction,
albeit with a small transverse angle, as in Fig.\ref{orbdet}b, due
to $\psi_{ingoing}$, which still has some (small) influence in that
region of the channel. Then there are two possibilities: i) the
trajectory traverses the whole channel and exits from it from the
side opposite to the entry, regaining almost horizontal flow
afterwards (Fig.\ref{orbdet}b), or ii) the orbit is entrained by the
channel all along its length, in which case the particle exits from
the scattering domain to infinity along the Bragg angle associated
with that particular channel (Fig.\ref{orbdet}c).
%_____________________________________________________________
%_____________________________________________________________

Since $\psi_{ingoing}$ is larger closer to the z-axis, the orbits
that started closer to the z-axis have larger probability to cross
more channels and end either with horizontal motion, or radial
motion along the direction of one of the first Bragg angles. This
leads to a complete stratification of the flow as shown in
Fig.\ref{orbits}. On the other hand, due to attenuation effects
(section 2) the channels practically disappear beyond some Bragg
angle and the separator takes locally the form of
Fig.\ref{channel}d, with oscillations of smaller and smaller
amplitude. This causes a gradually smaller concentration of the
exiting trajectories around the Bragg angles $\theta_q$ with high
$q$ , leading eventually to a diffuse form of the radial outward
flow at large angles $\theta$.

%%%%%%%%%%%%%%%%%%%%%%%%%%%%%%%%%%%%%%%%%%%%%%%%%%%%%%%%%%%%%%%%%%%%%
\section{Times of flight}
%%%%%%%%%%%%%%%%%%%%%%%%%%%%%%%%%%%%%%%%%%%%%%%%%%%%%%%%%%%%%%%%%%%%%
An important practical utility of the quantum trajectory approach
regards the possibility to unambiguously determine the {\it times of
flight} of particles, i.e. the time it takes for a particle to
travel between an emitter surface and a detector surface. This
question is of particular interest, because it is related to a well
known open problem of quantum theory, namely the so-called `problem
of time' (see \cite{muglea2000,mugetal2002} for reviews). This
problem stems from a theorem of Pauli \cite{pau1926}, according to
which it is not possible to properly define a self-adjoint time
operator consistent with all axioms of quantum mechanics. This
implies that the usual (Copenhagen)) formalism based on state
vectors or density matrices is not applicable to a
quantum-theoretical calculation of probabilities related to time
observables (e.g. the distribution of arrival times to a detector or
the times of flight defined as above). In fact, in standard quantum
mechanics (i.e. in both Schr\"{o}dinger's and Heisenberg's
pictures), time is considered only as a parameter of the quantum
equations of motion.

Among various proposals in the literature aiming to remedy this gap
of standard quantum theory (\cite{muglea2000}), the Bohmian
formalism offers a straightforward solution, since the time of
flight is well-defined along the Bohmian trajectories. This is in
principle subject to experimental testing, and one such test will be
proposed below. We note in passing two other approaches on the same
subject, namely the `history approach', based on Feynman paths, and
the Kijowski approach (\cite{kij1974}), based on so-called
`Bohm-Aharonov (\cite{hart1988,yata1993}) operators. However these
approaches have not been so far as clearly formulated as the Bohmian
formulation.

%_____________________________________________________________
%\begin{landscape}
\begin{figure}
\centering
\includegraphics[scale=0.50]{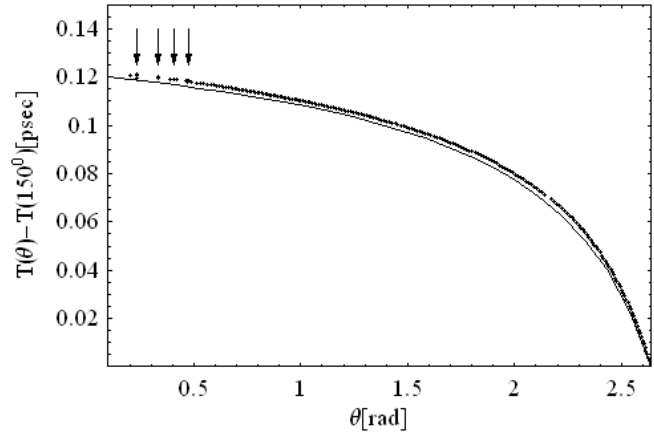}
\caption{The difference $T(\theta)-T(150^\circ)$ of the arrival
times of particles to detectors placed at the angles $\theta$ and
$150^\circ$ respectively and at the same distance from the center,
for the quantum trajectories of Fig.\ref{orbits}a. The solid curve
corresponds to the analytic estimate of Eq.(\ref{dtth2}). The arrows
correspond to the first four Bragg angles.} \label{plttheta}
\end{figure}
%\end{landscape}
%_____________________________________________________________
A calculation of the times of flight in the case of the trajectories
of Fig.\ref{orbits} can be done as follows. We note first that, as a
simple visual inspection of of Fig.\ref{orbits} shows, starting from
a fixed horizontal distance from the left (source), the trajectories
that are deflected to large angles are shorter in length than the
trajectories deflected to small angles, provided that the final
detection point is at the same distance $r$ from the center. We can
quantify this difference by approximating the form of the
trajectories as horizontal up to their deflection at the outer
separator and radial afterwards. The modulus of the velocity remains
nearly constant $v\simeq \hbar k_0/m$ in both the horizontal and
radial segments of the trajectory. Taking into account the separator
equation (Eq.(\ref{rthnod})), we estimate the length, and hence the
time difference between two trajectories deflected at two arbitrary
angles $\theta_1$ and $\theta_2$ by a simple geometric analysis. If
the separator is approximated by a straight line between the two
angles (e.g. in the forward and backward directions
$\theta_1=30^\circ$, $\theta_2=150^\circ$) we find:
\begin{equation}\label{dtth2}
T(\theta_1)-T(\theta_2)\approx {m\over\hbar k_0}
\left({(R_{0,2}+\lambda(\theta_1-\theta_2)(\cos\theta_1-1)\over\sin\theta_1}
-{R_{0,2}(\cos\theta_2-1)\over\sin\theta_2}\right)
\end{equation}
where the slope $\lambda$ is normalized to its value for the pair
$(\theta_1,\theta_2)=(30^\circ,150^\circ)$ namely
$$
\lambda={R_{0,2}-R_{0,1}\over \theta_2-\theta_1}
$$
with $R_{0,j}$, $j=1,2$ calculated by  Eq.(\ref{rthnod}) with
$\theta$ substituted by $\theta_1$ and $\theta_2$ respectively.

Eq.(\ref{dtth2}) agrees well with a numerical computation of the
times of flight for various angles $\theta$ (see
Fig.\ref{plttheta}). The main remark is that, since $R_0 \propto D$
(cf. Eq.(\ref{rthnod})), the time difference
$T(\theta_1)-T(\theta_2)$ for two fixed angles is proportional to
the transverse quantum coherence length $D$. That is,
Eq.(\ref{dtth2}) leads to the estimate
$T(\theta_1)-T(\theta_2)=O(D/v_0)$. This is a relation that can be
tested experimentally. In fact, it is very different to what is
found in the case of Rutherford scattering (compare
Figs.\ref{trjsch2}a and \ref{trjsch2}b), where a straightforward
analysis yields
\begin{eqnarray}\label{tthruth}
T(\theta_1)-T(\theta_2)\approx {Z Z_1 e^2\over 2\pi\epsilon_0m
v_0^3} \ln\left(\sqrt{1+\cot^2(\theta_2/2)\over
1+\cot^2(\theta_1/2)}\right)
\end{eqnarray}
i.e. the time difference $T(\theta_1)-T(\theta_2)$ does not depend
on the transverse quantum coherence length $D$ and it has a
different dependence $O(1/v_0^3)$, rather than $O(D/v_0)$, on the
particles' velocities. In the case of electrons, the classical time
difference is of order $10^{-20}sec$, while, assuming
$D\sim10^{-6}m$ the quantum time difference is of order
$10^{-12}sec$. We note that the latter time-resolution scale is
within present-day experimental possibilities. However, in order
that the experiment becomes feasible, one should be able to combine
very accurate time measurements with single-electron detectors (see
\cite{stei2008} for a review of recent experimental techniques on
quantum-mechanical time measurements). Furthermore, one should also
be able to have a control signal for the {\it emission} times of
electrons. This is a largely unexplored subject (one possibility is
probably offered by the so-called LASER induced field emission; see
\cite{baretal2007}). At any rate, in view of theoretical results
like the above, we think it is safe to anticipate that the advent of
experimental {\it time measurement} techniques in systems with a
genuinely quantum behavior will open a new window for probing
quantum mechanics at a very fundamental level.

%%%%%%%%%%%%%%%%%%%%%%%%%%%%%%%%%%%%%%%%%%%%%%%%%%%%%%%%%%%%%%%%%%%%%%%%%
\section{Conclusions}
%%%%%%%%%%%%%%%%%%%%%%%%%%%%%%%%%%%%%%%%%%%%%%%%%%%%%%%%%%%%%%%%%%%%%%%%%
We applied the method of the de Broglie - Bohm quantum trajectories
in the problem of charged particle diffraction from thin material
targets, focusing on the case of significant transverse quantum
coherence of the particle beam, where new genuinely quantum
phenomena appear. In
particular: \\

1) We constructed a model for the wavefunction of diffracted
particles which takes into account both processes of coherent
(giving rise to Bragg angles) and diffuse scattering.

2) We developed a theory for the quantum-current structure near a
locus called {\it separator}, i.e. the border between an inner tube
of ingoing flow surrounding the beam's axis of symmetry and an outer
domain, where the radial outward flow prevails. Analytical
expressions are found for the separator. The separator forms thin
channels of radial flow very close to every Bragg angle. Such
channels are responsible for the concentration of quantum
trajectories to particular directions of exit from the ingoing flux
tube.

3) The deflection of quantum trajectories is due to their
interaction with an array of {\it quantum vortices} formed around a
large number of nodal points located on the separator. The quantum
flow near every nodal point takes the form of a `nodal point -
X-point complex'. The size of quantum vortices is estimated
analytically close to and far from Bragg angles, the estimate being
in close agreement with numerical results.

4) The emergence of a diffraction pattern is explained in terms of
numerically calculated quantum trajectories. In particular, we
demonstrate the sharp deflections of the trajectories as they
approach one or more X-points along the latters' stable manifolds,
and recede from these points along their unstable manifolds. The
radically non-classical character of the quantum trajectories is
demonstrated. We find that trajectories with larger initial distance
$R_0$ from the central axis of the ingoing flux tube (i.e. larger
impact parameter) are deflected to larger angles $\theta$. This is
contrary to the classical Rutherford scattering, where the
trajectories with larger impact parameter are deflected to smaller
angles $\theta$.

5) The times of flight $T(\theta)$ of particles to detectors placed
at constant distances  and various angles $\theta$ from the target
are calculated by an analytical approximation and compared to
numerical results. It is demonstrated that the time difference
$T(\theta_1)-T(\theta_2)$ follows the scaling
$T(\theta_1)-T(\theta_2)=O(D/v_0)$, where $D$ is the transverse
quantum coherence length and $v_0$ the average particles' velocity.
This scaling is different than in classical Rutherford scattering
where $T(\theta_1)-T(\theta_2)=O(|Z Z_1|
e^2/(2\pi\epsilon_0mv_0^3))$ and it leads to the possibility of an
experimental test of the Bohmian formalism in a sector of quantum
theory where the Copenhagen approach offers no standard recipe, i.e.
the sector of time observables.
 \nonumsection{Acknowledgments} \noindent C. Delis was
supported by the State Scholarship Foundation of Greece (IKY) and by
the Hellenic Center of Metals Research.

\appendix{\textbf{Effective Fraunhofer function}}

We consider the simplest example of a cubic lattice arrangement of
the atoms in the target. The atomic positions $\mathbf{r}_j$ are
given by:
\begin{eqnarray}\label{lattice}
&~&\mathbf{r}_j=(n_x,n_y,n_z)a + \Delta a\mathbf{u}_j(t), \nonumber\\
&~&\\
(n_x,n_y,n_z)&\in& (-{N_\perp\over 2},{N_\perp\over 2})\times
(-{N_\perp\over 2},{N_\perp\over 2})\times (-{N_z\over 2},{N_z\over
2})\nonumber
\end{eqnarray}
where $a$ is the lattice constant (equal to the length of one side
of the primitive cell), $\Delta a$ is the amplitude of random
oscillations (due to thermal or recoil motions; $\Delta a$ is taken
equal to a small fraction of $a$) and
$\mathbf{u_j}\equiv(u_{j,x},u_{j,y},u_{j,z})$ are random variables
with a uniform distribution in the intervals $[-0.5,0.5]$. The
number of atoms $N_z$ in the z-direction is $N_z=d/a$, where $d$ is
the target thickness. On the other hand, the value of $N_\perp$ can
be chosen as $N_\perp=D/a$, since by such a choice the value of the
Gaussian weight in Eq.(\ref{seff}) can be approximated by $\approx
1$ for all $|n_x|<N_\perp/2$ and $|n_y|<N_\perp/2$, and $\approx 0$
for $|n_x|>N_\perp/2$ or $|n_y|>N_\perp/2$. Given these
approximations, we find a model for the effective Fraunhofer
function (Eq.(\ref{seff})) accounting for diffraction effects when
$k_0a\sim 1$ (if $k_0a>>1$ the diffraction effects disappear). In
this case we have
\begin{eqnarray}\label{seff2}
S_{eff}(k_0;\mathbf{r})\approx
\sum_{n_z=-N_z/2}^{N_z/2}\sum_{n_x=-N_\perp/2}^{N_\perp/2}
\sum_{n_y=-N_\perp/2}^{N_\perp/2}
e^{2ik_0(n_za+\Delta a u_{z})\sin^2(\theta/2)}\nonumber\\
\times  e^{-i[(n_xa+\Delta a
u_{x})\sin\theta\cos\varphi-(n_ya+\Delta a
u_{y})\sin\theta\sin\varphi]}
\end{eqnarray}
Equation (\ref{seff2}) is still not convenient for a practical use
in numerical calculations, since the triple sum in the r.h.s. has
a prohibitive cost to compute. However, a drastic simplification
takes place by considering a material target having a polycrystalline
structure. In this case, we can effectively proceed by randomizing the
value of $\varphi$ in the terms of (\ref{seff2}) (which is mathematically
equivalent to considering random rotations of small crystallites, on
planes normal to the beam). In this case, the dependence of the sum in
Eq.(\ref{seff2}) on $u_x$, $u_y$ effectively disappears, since
the corresponding terms in the exponential argument are effectively
dominated by the random variations due to $\varphi$. Furthermore,
since $u_z$ is also random, the distribution of values of $S_{eff}$
for different samples of values $(u_x,u_y,u_z)$ becomes practically
equivalent to the distribution of values of the quantity
\begin{eqnarray}\label{seff3}
S_{eff}(k_0;\mathbf{r})\approx
\sum_{n_z=-N_z/2}^{N_z/2}e^{2ik_0(n_za+\Delta a u_{z})\sin^2(\theta/2)}\nonumber\\
\times \sum_{n_x=-N_\perp/2}^{N_\perp/2}
\sum_{n_y=-N_\perp/2}^{N_\perp/2} e^{-i[(n_xa+\Delta a
u_{x})\sin\theta\cos\varphi-(n_ya+\Delta a
u_{y})\sin\theta\sin\varphi]}
\end{eqnarray}
i.e. where the triple sum is decomposed to a $1\times 2$ sum.
Denoting by $S_{xy}$ an rms value of the double sum in (\ref{seff3}),
we find that $S_{xy}$ is of order $N_\perp\sim D/a$. We then substitute
$S_{xy}=(D/a) C_{coherent}$, where $C_{coherent}$ is a fitting
constant, in the place of the double sum.  Equation (\ref{seff3}) takes
the form
\begin{eqnarray}\label{seffz}
S_{eff}(\theta)\approx
(D/a)e^{i\delta}\sum_{n_z=-N_z/2}^{N_z/2}C_{coherent}\exp(i2k_0(n_za+\Delta
au_{n_z})\sin^2(\theta/2))
\end{eqnarray}
where $\delta$ is a random phase. The value of the fitting constant
$C_{coherent}$ is determined numerically by running a realization of
the sums appearing in (\ref{seffz}) in the computer, and comparing
the computed quantities with the results with the full sum
(\ref{seff2}) for various values of $N_\perp$ and $N_z$. To arrive
at a final fitting model, we note first that if the noise is taken
equal to zero ($u_{n_z}=0$), the form of $S_{eff}$ near an angle
$\theta_q$ is:
\begin{equation}\label{seffbragg}
S_{q,eff}(\theta)\approx (D/a)e^{i\delta} C_{coherent}
\frac{2\sin\left[k_0N_za\sin(\theta_q)(\theta-\theta_q)/2\right]}
{k_0a\sin(\theta_q)(\theta-\theta_q)}~~.
\end{equation}
Taking now into account the noise, the attenuation of the maxima can
be estimated by a Debye-Waller factor (see e.g. \cite{peng2005}),
changing (\ref{seffbragg}) into
\begin{eqnarray}\label{seffdw}
& &S_{q,eff}(\theta)\approx\nonumber\\
& &(D/a)e^{i\delta}  C_{coherent}\exp\left(-{1\over
2}4k_0^2\sin^4(\theta/2)\sigma_a^2\right)
\frac{2\sin\left[k_0N_za\sin(\theta_q)(\theta-\theta_q)/2\right]}
{k_0a\sin(\theta_q)(\theta-\theta_q)}
\end{eqnarray}
Equation (\ref{seffdw}) yields the form of the Fraunhofer function
locally, very close to a Bragg angle. The total `coherent'
contribution to the Fraunhofer function is a sum of terms like
(\ref{seffdw}) over all Bragg angles:
\begin{eqnarray}\label{seffdw2}
& &S_{coherent}(\theta)\approx\nonumber\\\
& &(D/a)e^{i\delta} \sum_{q=0}^{q_{max}}
C_{coherent}\exp\left(-{1\over
2}4k_0^2\sin^4(\theta/2)\sigma_a^2\right)
\frac{2\sin\left[k_0N_za\sin(\theta_q)(\theta-\theta_b)/2\right]}
{k_0a\sin(\theta_q)(\theta-\theta_q)}~~
\end{eqnarray}
where $q_{max}$ is the total number of Bragg angles. To this we add
a `diffuse' term accounting for random phasor sums away from all
Bragg angles. We have $S_{diffuse}\sim N_\perp N_z^{1/2}$, weighted
by the complement of the Debye-Waller factor $1-e^{-{1\over
2}4k_0^2\sin^4(\theta/2)\sigma_a^2}$. Thus, a final model for the
Fraunhofer function reads:
\begin{eqnarray}\label{sbfin2}
S_{eff}(\theta)&=&{De^{i\delta}\over a }\Bigg[ \sum_{q=0}^{q_{max}}
C_{coherent}e^{-{1\over 2}4k_0^2\sin^4(\theta/2)\sigma_a^2}
\frac{2\sin\left[k_0N_za\sin(\theta_q)(\theta-\theta_q)/2\right]}
{k_0a\sin(\theta_q)(\theta-\theta_q)}\nonumber\\
&+&(1-e^{-{1\over
2}4k_0^2\sin^4(\theta/2)\sigma_a^2})C_{diffuse}\sqrt{N_z}\Bigg]
\end{eqnarray}
where $C_{diffuse}$ is also a fitting constant specified
numerically. Substituting $N_z=d/a$ we obtain Eq.(\ref{sbfin}).

\newpage

\appendix{\textbf{Size of quantum vortices}}

Starting from Eq.(\ref{psiexp}), we obtain an estimate of the size
of the coefficients $a_{ij}$, $b_{ij}$ near and far from Bragg
angles as follows: For an angle $\theta_0$ close to a Bragg angle,
using the definition (\ref{seffbragg}), one has for the derivatives
of $S_{eff}(\theta)$ the following estimates:
$$
{d S_{eff}\over S_{eff}d\theta_0}\sim {N_z k_0 a} \sim k_0 d,~~~
{d^2S_{eff}\over S_{eff}d\theta_0^2}\sim {N_z^2 k_0^2 a^2} \sim (k_0
d)^2~~.
$$
Since $d/D<<1$, from the above expressions we find the dominant
terms of all the coefficients $a_{ij}$, $b_{ij}$. These are:
\begin{eqnarray}\label{coefdoma2}
&a_{10}=e^{-R_0^2/2D^2}\bigg[-\sin(k_0z_0) k_0 (1-\cos\theta_0)+ O(k_0 d/D)\bigg]\nonumber\\
&b_{10}=e^{-R_0^2/2D^2}\bigg[\cos(k_0z_0) k_0 (1-\cos\theta_0)+ O(k_0 d/D)\bigg]\\
&a_{01}=e^{-R_0^2/2D^2}\bigg[\sin(k_0z_0) k_0 \sin\theta_0+ O(k_0 d/D)\bigg]\nonumber\\
&b_{01}=e^{-R_0^2/2D^2}\bigg[-\cos(k_0z_0) k_0 \sin\theta_0+ O(k_0
d/D)\bigg]\nonumber
\end{eqnarray}
\begin{eqnarray}\label{coefdoma2}
&a_{20}=e^{-R_0^2/2D^2}\bigg[-\cos(k_0z_0) {k_0^2\over 2} \sin^2\theta_0 + O(k_0^2 d^2/D^2)\bigg]\nonumber\\
&b_{20}=e^{-R_0^2/2D^2}\bigg[-\sin(k_0z_0) {k_0^2\over 2} \sin^2\theta_0 + O(k_0^2 d^2/D^2)\bigg]\nonumber\\
&a_{02}=e^{-R_0^2/2D^2}\bigg[\cos(k_0z_0) {k_0^2\over 2} \sin^2\theta_0 + O(k_0^2 d^2/D^2)\bigg]\\
&b_{02}=e^{-R_0^2/2D^2}\bigg[\sin(k_0z_0) {k_0^2\over 2} \sin^2\theta_0 + O(k_0^2 d^2/D^2)\bigg]\nonumber\\
&a_{11}=e^{-R_0^2/2D^2}\bigg[\cos(k_0z_0) {k_0^2\over 2} \sin 2\theta_0 + O(k_0^2 d^2/D^2)\bigg]\nonumber\\
&b_{11}=e^{-R_0^2/2D^2}\bigg[\sin(k_0z_0) {k_0^2\over 2} \sin
2\theta_0 + O(k_0^2 d^2/D^2)\bigg]\nonumber
\end{eqnarray}
The quantities $r_0,z_0,R_0$ are all of order $O(D)$. The equalities
$a_{20}=-a_{02}$, and $b_{20}=-b_{02}$ (to leading order) are due to
the fulfilment of the continuity equation by the wavefunction $\psi$
(see \cite{eftetal2009}). However, since the outgoing term was
evaluated only up to $O(\sigma_\perp^2/k_0^2)=O(1/D^2k_0^2)$, the
above equalities are also violated at the same order, resulting in a
small relative error (of order $10^{-6}$). Substituting these
expressions into Eq.(\ref{xpointa2}), we remark that the products of
the leading terms of the coefficients $a_{10},b_{10},a_{01},b_{01}$
{\it cancel exactly in the coefficient $A$} (while they do not
cancel in all other contributions). As a result, near a Bragg angle
one has
$$
A=O\left({k_0^2 d\over
D}\right),~~B_i=O(k_0^3),~~C_i=O(k_0^3),~~D_i=O(k_0^3),~~i=1,2
$$
whence, in view of (\ref{xpointa2}),
\begin{equation}\label{xpesta2}
u_X=O\left({d\over D k_0}\right),~~~~v_X=O\left({d\over D
k_0}\right)~~.
\end{equation}
Substituting the above estimates in the expression
$R_X=(u_X^2+v_X^2)^{1/2}$ we find the first of Eqs.(\ref{rxest}).

On the other hand, far from Bragg angles all the terms with
derivatives $dS_{eff}/d\theta_0$, $d^2S_{eff}/d\theta_0^2$ in the
equations specifying the coefficients $a_{ij},b_{ij}$ become
negligible. Then, we have precisely the same leading terms in all
the coefficients $a_{ij},b_{ij}$ as before. In the subsequent order,
however, the terms are $O(1/D)$ for the coefficients
$a_{10},b_{10}$, $a_{01},b_{01}$, and $O(k_0/D)$ for the
coefficients $a_{20},b_{20}$, $a_{11},b_{11}$, $a_{02},b_{02}$. Then
$$
A=O\left({k_0\over
D}\right),~~B_i=O(k_0^3),~~C_i=O(k_0^3),~~D_i=O(k_0^3),~~i=1,2
$$
Thus
\begin{equation}\label{xpnesta2}
u_X=O\left({1\over D k_0^2}\right),~~~~v_X=O\left({1\over D
k_0^2}\right)~~
\end{equation}
which, upon substitution to $R_X=(u_X^2+v_X^2)^{1/2}$ leads to the
second of the estimates (\ref{rxest}).

%%%%%%%%%%%%%%%%%%%%%%%%%%%%%%%%%%%%%%%%%%%%%%%%%%%%%%%%%%%%%%%%%%%%%%%%%%%%%%%%%%%%%%
%\section*{References}
%%%%%%%%%%%%%%%%%%%%%%%%%%%%%%%%%%%%%%%%%%%%%%%%%%%%%%%%%%%%%%%%%%%%%%%%%%%%%%%%%%%%%%

%\end{multicols}

\begin{thebibliography}{9}

\bibitem[Barwick {\it et al.}(2007)]{baretal2007}
Barwick, B., Corder, C., Strohaber, J., Chandler-Smith, N.,
Uiterwaal, C., \& Batelaan, H.\ 2007, New Journal of Physics, 9, 142

\bibitem[Beenakker \& van Houten(1991)]{beehou1991}
Beenakker, C.W., and van Houten, H.: 1991, {\it Solid State Phys.}
{\bf 44}, 1.

\bibitem[Bennett(2010)]{ben2010}
Bennett, A.: 2010, {\it J. Phys. A} {\bf 43}, 5304.

\bibitem[Bohm(1952)]{bohm1952}
Bohm, D.: 1952,{\it Phys. Rev.} {\bf 85}, 166; {\bf 85} 194.

\bibitem[Colin \& Struyve(2010)]{colstru2010}
Colin, S., and Struyve, W.: 2010, {\it New J. Phys.} {\bf 12}, 3008.

\bibitem[Contopoulos \& Efthymiopoulos(2008)]{coneft2008}
Contopoulos, G., and Efthymiopoulos, C.:2008, {\it Celest. Mech.
Dyn. Astron.} {\bf 102}, 219.

\bibitem[De Broglie(1925)]{debro1925}
De Broglie, L.: 1925,{\it Ann. Phys. Paris} {\bf 3}, 22.

\bibitem[Delis {\it et al.}(2011)]{deletal2011}
Delis, N., Efthymiopoulos C., and Contopoulos, G.: 2011,
`Wavepacket approach to electron diffraction: Bohmian trajectories
and arrival times' (in preparation).

\bibitem[Efthymiopoulos \& and Contopoulos(2006)]{eftcon2006}
Efthymiopoulos C., and Contopoulos, G.: 2006, {\it J. Phys. A} {\bf
39}, 1819.

\bibitem[Efthymiopoulos {\it et al.}(2007)]{eftetal2007}
Efthymiopoulos, C., Kalapotharakos, C., and Contopoulos, G.: 2007,
{\it J. Phys. A} {\bf 40}, 12945.

\bibitem[Efthymiopoulos {\it et al.}(2009)]{eftetal2009}
Efthymiopoulos, C., Kalapotharakos, C., and Contopoulos, G.: 2009,
{Phys. Rev. E} {\bf 79}, 036203.

\bibitem[Frisk(1997)]{fri1997}
Frisk, H.: 1997, {\it Phys. Lett. A} {\bf 227}, 139.

\bibitem[Gindensperger(2003)]{gin2003}
Gindensperger, E.: 2003, {\it Phys. Rev. } {\bf D37}, 2818.

\bibitem[Hartle(1988)]{hart1988}
Hartle, J.B.: 1988, {\it Phys. Rev. } {\bf D37}, 2818.

\bibitem[Kijowski(1974)]{kij1974}
Kijowski, J: 1974, {\it Rev. Mod. Phys.} {\bf 6}, 361.

\bibitem[Lopreore(1999)]{lopwya1999}
Lopreore, C.L. and Wyatt, R.E.: 1999, {\it Phys. Rev. Lett.} {\bf
82}, 5190.

\bibitem[Madelung(1926)]{mad1926}
Madelung, E: 1926,{\it Z. Phys.} {\bf 40}, 332.

\bibitem[Muga \& Leavens(2000)]{muglea2000}
Muga, J.G., and Leavens, C.R.: 2000, {\it Phys. Rep.} {\bf 338},
353.

\bibitem[Muga {\it et al.}(2002)]{mugetal2002}
Muga, J.G., Sala Mayato, R., and Egusquiza I.L.: 2002, {\it Lect.
Notes Phys.} {\bf 72}, 1.

\bibitem[Oriols(2007)]{ori2007}
Oriols, X: 2007, {\it Phys. Rev. Lett.} {\bf 98}, 066803.

\bibitem[Pauli(1926)]{pau1926}
Pauli, W: 1926, {\it Hanbuch der Physik} {\bf 22}, pp.1-278, Springer,
Berlin.

\bibitem[Peng {\it et al.}(2004)]{pengetal2004}
Peng, L.M., Dudarev, S.L., and Whelan, M.J.: 2004, {\it High-Energy
Electron Diffraction and Microscopy}, Oxford University Press,
Oxford.

\bibitem[Peng(2005)]{peng2005}
Peng, L.M.: 2005, {\it J. El. Micr.} {\bf 54}, 199.

\bibitem[Philippidis {\it et al.}(1979)]{phietal1979}
Philippidis, C., Dewdney, C. and Hiley, B.: 1979, {\it Nuovo Cimento
B} {\bf 52}, 15.

\bibitem[Sanz {\it et al.}(2002)]{sanzetal2002}
Sanz, A.S., Borondo, F., and Miret-Art\'{e}s, S: 2002, {\it J.
Phys.: Condens. Matter} {\bf 14}, 6109.

\bibitem[Sanz {\it et al.}(2004)]{sanzetal2004a}
Sanz,A.S., Borondo, F., and Miret-Art\'{e}s, S: 2004, {\it J. Chem.
Phys.} {\bf 120}, 8794.

\bibitem[Steinberg(2008)]{stei2008}
Steinberg, A.S.: 2008, {\it Lect. Notes Phys.} {\bf 734}, 333.

\bibitem[Vallentini \& and Westman(2005)]{valwes2005}
Vallentini, A., and Westman, H.: 2005, {\it Royal Soc. London Proc.
A} {\bf 461}, 253.

\bibitem[Wyatt(2005)]{wya2005}
Wyatt, R.: 2005,{\it Quantum Dynamics with Trajectories}, Springer,
New York.

\bibitem[Wisniacki \& Pujals(2005)]{wispuj2005}
Wisniacki, D.A., and Pujals, E.R.: 2005, {\it Europhys. Lett.} {\bf
71}, 159.

\bibitem[Yamada \& Takagi(1993)]{yata1993}
Yamada, N. and Takagi, S.: 1993, {\it Prog. Theor. Phys.} {\bf 86},
599.

\bibitem[Zhao \& Makri(2003)]{zhamak2003}
Zhao, Y., and Makri, N.: 2003, {\it J. Chem. Phys.} {\bf 119}, 60.

\end{thebibliography}
\end{document}